\documentstyle[twoside,fleqn,espcrc2,epsf]{article}
\newcommand{\be}{\begin{equation}}
\newcommand{\ee}{\end{equation}}
\title{Flavour-singlet pseudoscalar and scalar mesons }
\author{UKQCD Collaboration, 
        C.~Michael, M. S. Foster\thanks{present address: MVC, Manchester
Computing, Manchester M13 9PL} and C. McNeile \\ [8pt] 
Theoretical Physics Division, Dept. of Mathematical Sciences, 
          University of Liverpool, Liverpool L69 3BX, UK }
\begin{document}
%%%%%%%%%%%%%%%%%%%%%%%%%%%%%%%%%%%%%%%%%%%%%%%%%%%%%%%%%%%%%%%%%%
\begin{abstract}
 
We measure correlations appropriate for favour-singlet mesons using 
dynamical quark configurations from UKQCD. Improved methods of 
evaluating the disconnected quark diagrams are presented. The  scalar
and pseudoscalar meson channels are explored, with special reference to
the  mixing of scalar mesons with scalar glueballs.

\end{abstract}
\maketitle

%%%%%%%%%%%%%%%%%%%%%%%%%%%%%%%%%%%%%%%%%%%%%%%%%%%%%%%%%%%%%%%%%%

\section{INTRODUCTION}

 To study flavour singlet mesons, we need to consider quark loops  which
are disconnected (often called hairpins), namely evaluate $\rm{Tr}
\Gamma M^{-1}$ where the sum (trace) is over all space (for zero
momentum) at a given  time value and all colours and spins. Here $M$ is
the lattice fermion matrix, $\Gamma$ is a combination of the 
appropriate $\gamma$-matrix and a product of gauge links if a non-local
operator is used for the meson.

 Using a random volume source $\xi$ (where $\langle \xi^* \xi \rangle=1$
 for the same colour, spin and space-time component and zero otherwise)
then  solving $M \phi = \xi$, one can  evaluate unbiased estimates of
the propagator $M^{-1}_{xy}$  from $\langle \xi^*_y \phi_x \rangle$
where the average is over the $N_S$ samples  of the stochastic source.
The drawback of this approach is that
the variance on these estimates can be very large, so that typically 
hundreds of samples are needed.  Here we present a method which succeeds
 in reducing this variance substantially at rather small computational
expense.

 The variance reduction  is based on expressing the fermion
matrix $M$ as
 \be
 M=C+D = C(1+C^{-1}D)=(1+DC^{-1})C
 \ee
 where $C$ is easy to invert, for example the SW-clover term which is local 
in space. Then we have the exact identity 
 \begin{eqnarray}
  M^{-1} \!\!\!\!\! & = & \!\!\!\!C^{-1}  -C^{-1}D C^{-1} + \dots + 
               (-C^{-1}D)^m C^{-1} \nonumber \\
 & +&  (-C^{-1} D)^{n_1} M^{-1}(-D C^{-1})^{n_2}  
  \end{eqnarray}
      with $n_1+n_2=n=m+1$.

 Using the stochastic estimate for $M^{-1}$ on the rhs will reduce the 
variance of the estimate given by the lhs since the terms not involving 
$M^{-1}$ can be evaluated either exactly (for example terms with odd
powers of $D$ vanish in the evaluation  of a local trace) or as a
subsidiary  stochastic calculation with more samples since no inversion
is required.

 A special case of this ($n_1=n_2=2$) with Wilson fermions (for which $C=1$ 
and the terms with up to 3 powers of $D$ vanish for $\rm{Tr} M^{-1}$)
was used by the bermion group~\cite{bermions} previously.

 Using larger values of $n_1$ and $n_2$ implies that the estimate of 
$M^{-1}$ is very non-local. To evaluate correlators between traces  at
$t_1$ and $t_2$, one must require that the samples of stochastic  volume
source used in the two cases are different so that there is no bias.  We
use $N_S=24$ stochastic samples and this condition is readily
implemented.  This number of samples was chosen to make the stochastic
sampling error smaller (actually about 50\% in the cases to be discussed
next) than  the intrinsic variance from one gauge configuration to
another.  The computational effort  is equivalent to that in obtaining 
two conventional propagators (from two sources of  all colour-spins)

 We now apply this technique to  12$^3$ 24 lattices with $N_f=2$ at
$\beta=5.2$ with $C_{SW}=1.76$ from UKQCD~\cite{ukqcd}. We use valence
quark masses equal to the  sea-quark mass determined by the
$\kappa$-values as given in Table~1. In order to improve the statistics
we measure the disconnected diagrams on  configurations separated by 
less trajectories than for the connected correlators giving  253, 169
configurations  at $\kappa=0.1395$, 0.1398, respectively.

\section{PSEUDOSCALAR MESONS}

 Since we are exploring $N_f=2$ we describe the flavour non-singlet 
($I=1$) pseudoscalar meson as $\pi$ and the flavour singlet ($I=0$)  as
$\eta$. It is possible to apply our results to the physical case  of 3
light quark flavours and the $\eta, \eta'$ splitting as has been  done
from the quenched case~\cite{kuramashi}, but we do not discuss this
here.

 We use local and fuzzed operators matching the calculation of the
connected  meson propagators~\cite{ukqcd}. Then including our results
for  the hairpin diagrams with the appropriate sign, we can fit the
flavour singlet correlator to the  $\eta$ mass and excited states. Even
though, as discussed above, we have  measured the disconnected quark
loop correlator from every time slice to every  other with a stochastic
error which is effectively negligible, the signal  is very noisy. This
arises since the disconnected part of the  $\eta$
correlation has an error which does not decrease with increasing  time
separation $T$, unlike the connected correlator where the  error is
roughly constant as a percentage of the signal with increasing $t$.  To
illustrate this, we show the ratio  of disconnected $D$ to connected
component $C$ of the $\eta$ correlator in fig.~1.

The line shows the expectation coming from  one state ($\pi$)
contributing  to $C$ and one state ($\eta$) contributing to $C-D$ with  a
mass difference in lattice units of 0.16. There is some sign in  our
data of the impact of unitarity for matched valence and sea quarks:
namely that $D/C$ approaches  1 from below at large $t$. Making a fit
with this constraint to $D/C$,  it is possible to estimate the $\eta -
\pi$ mass difference and we see that  it does increase as the sea quark
mass is decreased.  Assuming that the $\eta$ mass is  constant as the
quark mass is decreased, we would get an $\eta$ mass  in the chiral
limit of around $800$ MeV, of course with an uncontrolled systematic
error

Because the flavour singlet correlator  becomes noisy so rapidly with
increasing time separation, it is  very valuable to use a more extended
basis for the  meson (eg. local and fuzzed operators) to help in
extracting the ground state signal. The 2 state fit from $t=3$ to 9 is
shown in Table~1.

The zero-momentum correlator has contributions increasing like  $L^3$
for spatial volume $L^3$ whereas the noise increases like $L^6$. Thus we
 find a better signal to noise ratio for smaller volumes - hence we
present here only  the results for $L=12$.

\begin{figure}[htb]
\vspace{-3.3cm}
\epsfxsize=9.5cm\epsfbox{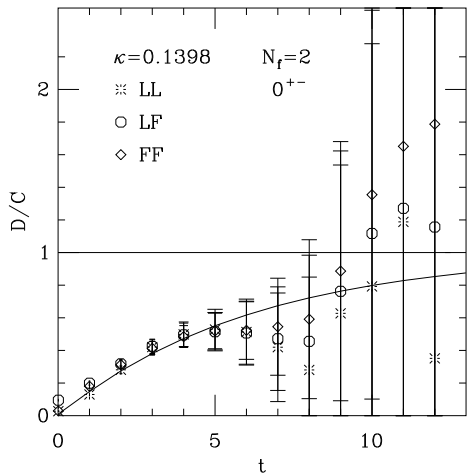}
\vskip -1.7cm
\hskip -6mm
 \caption{Disconnected to connected ratio. }
 \label{dbyc}
\vspace{-0.7cm}
\end{figure}

\begin{table}[t]
\footnotesize
\begin{tabular}{lllll}
\hline
 $\kappa$ & $r_0/a$ & $m_{\pi}a$ & $m_{\pi}/m_{\rho}$ & $m_{\eta}a$ \\
\hline  
0.1395 & 3.44 & 0.558(8) & 0.71 &  0.65(6) \\
0.1398 & 3.65 & 0.476(16) & 0.67 & 0.62(15) \\
\hline

\end{tabular}
 \caption{}

 \label{masses}
\vspace{-12mm}
\end{table}

\section{SCALAR MESONS}

 For scalar mesons, we have the interesting situation that the  $q
\bar{q}$ mesons and the glueball can mix. Within the quenched
approximation,  it is possible to estimate this mixing~\cite{lw}. We
make a preliminary study in the quenched approximation at $\beta=5.7$
of 100 configurations on 12$^3$ 24 lattices  with SW-clover valence
quarks having $\kappa=0.14077$ with $C_{SW}=1.57$ (here $a \approx 1$ GeV
and the quark mass  is close to strange). We use 4 scalar meson
operators (i) closed Wilson loops (glueball operators) of  two
different sizes (Teper-smeared) and (ii) $q \bar{q}$ operators which are
 local and separated by fuzzed links. Using the glueball mass of 0.97(4)
from ref.\cite{lw} and scalar $q\bar{q}$ mass  of 1.48(15) from fitting
the  connected correlations, we are able to fit the disconnected
correlation  and the hairpin-Wilson loop correlation from $t$-values of
1 to 3  with a mixing given by $Ea=0.4$ assuming $N_f=2$. This fit (see 
fig.~2) assumes only ground state contributions so that  the systematic
error on $E$ from this assumption is hard to estimate. The mixing
estimated by ref~\cite{lw} is similar  in magnitude at these lattice
parameters ($E \approx 0.3$ GeV) but they claim that on extrapolation to
the continuum limit  a much smaller value is obtained. Here we are using
clover improvement so order $a$ effects are suppressed.  If our 
quenched mixing strength were be to applied to the scalar mass matrix,
it results in a downward shift for $N_f=2$ of the lattice glueball  mass
by 20\%.

\begin{figure}[htb]
\vspace{-3.3cm}
\epsfxsize=9.5cm\epsfbox{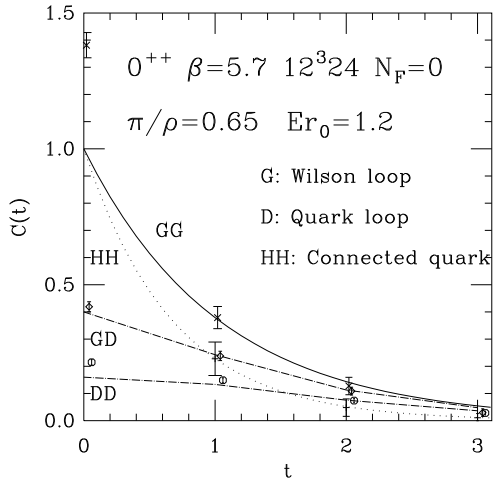}
\vskip -1.7cm
\hskip -6mm
 \caption{Quenched scalar correlations.
 }
 \label{mixing}
\vspace{-0.7cm}
\end{figure}

 We can now measure directly the scalar spectrum for $N_f=2$ to explore
this.  We again use both glueball and $q \bar{q}$ operators and fit with
2 states  the $4 \times 4$ matrix of correlations for $t$ from 2 to 10:
the results is shown as $m_{FS}a$  in Table 2. We find that the  mass
obtained from fitting only the glueball correlations ($m_{GB}$) is
consistent with the  full fit, as it should be. Moreover,  we see a
surprisingly low  scalar mass - as emphasised in fig.~3 which compares
with quenched results and the SESAM $N_f=2$ values~\cite{SESAM}.

\begin{table}[tb]
\footnotesize
\begin{tabular}{lllll}
\hline
 $\kappa$ & $m_{GB}a$ & $m_{FS}a$ & $m_{NS}a$ \\
\hline  
0.1395 & 0.47(6)&0.44(3)  & 1.28(7)\\ % 0.44(3) 2-10 fact fit, 0.47(6) GB
0.1398 & 0.50(10)&0.52(8)  &1.03(7) \\ % 0.52(8) 2-10 fact fit, 0.50(10) GB
\hline
\end{tabular}
\caption{}
 \label{scalar}
\vspace{-12mm}
\end{table}

We do expect a relatively light flavour-singlet scalar mass because of
mixing  effects as described above which would reduce the 
mass by 20\%. This could explain in part our low
scalar mass but  other explanations are also worth exploring. For
example the order $a^2$  corrections might be anomalously large for our
lattice implementation (e.g. twice as large as in the quenched
Wilson case).

 Another possible explanation of the light flavour-singlet scalar mass
we find (comparable to our pion mass) would be a partial restoration of
chiral symmetry  in a finite volume. Our spatial size  is 1.7
fm and no evidence of finite size effects  was seen in a study of
flavour non-singlet correlators~\cite{ukqcd}.
 We have made a preliminary study of flavour singlet correlators on 
$16^3$ spatial lattices to check for  finite size effects and, although 
the signal from the larger spatial volume is relatively noisier,  we do
see some evidence (at the $2\sigma$ level) of a  higher scalar mass on
the larger spatial lattice.

\begin{figure}[htb]
\vspace{-3.3cm}
\epsfxsize=9.5cm\epsfbox{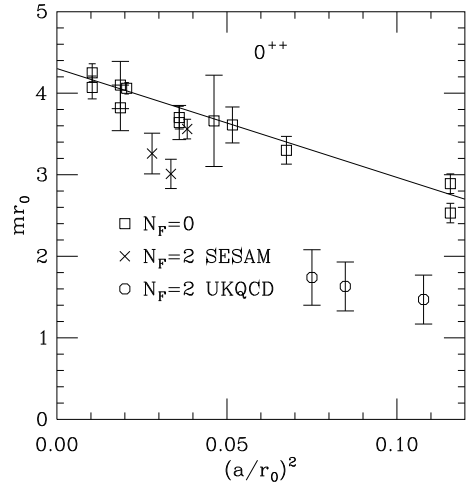}
\vskip -1.7cm
\hskip -6mm
 \caption{ The scalar mass versus $a^2$.
 }
 \label{gbr0}
\vspace{-0.7cm}
\end{figure}

\end{document}